\documentclass[twocolumn,notitlepage,superscriptaddress,floatfix]{revtex4-1}

\usepackage{hyperref}
\usepackage{mathtools}
\usepackage{graphicx}
\usepackage{xcolor}
\usepackage[export]{adjustbox}
\usepackage[normalem]{ulem}
\usepackage{braket}

\definecolor{p-channel}{RGB}{91,184,159}
\definecolor{c-channel}{RGB}{227,171,122}
\definecolor{d-channel}{RGB}{179,144,212}
\definecolor{vertex_fill}{RGB}{125,125,125}

\newcommand{\captiontitle}[1]{\textbf{#1}}
\newcommand{\bvec}[1]{\boldsymbol{#1}}

\newcommand{\delete}[1]{}
\newcommand{\append}[1]{#1}
\newcommand{\substitute}[2]{\delete{#1}\append{#2}}

\begin{document}
\title{Moir\'e Engineering of Spin-Orbit Coupling in \append{Twisted} Platinum Diselenide}

\author{Lennart Klebl}
\altaffiliation{L.K. and Q.X. contributed equally to this paper.}
\affiliation{Institut f\"ur Theorie der Statistischen Physik, RWTH Aachen University and JARA-Fundamentals of Future Information Technology, 52056 Aachen, Germany}

\author{Qiaoling Xu}
\altaffiliation{L.K. and Q.X. contributed equally to this paper.}
\affiliation{Songshan Lake Materials Laboratory, 523808 Dongguan, Guangdong, China}
\affiliation{College of Physics and Electronic Engineering, Center for Computational Sciences, Sichuan Normal University, Chengdu 610068, China}

\author{Ammon Fischer}
\affiliation{Institut f\"ur Theorie der Statistischen Physik, RWTH Aachen University and JARA-Fundamentals of Future Information Technology, 52056 Aachen, Germany}

\author{Lede Xian}
\altaffiliation{xianlede@sslab.org.cn}
\affiliation{Songshan Lake Materials Laboratory, 523808 Dongguan, Guangdong, China}
\affiliation{Max Planck Institute for the Structure and Dynamics of Matter, Center for Free Electron Laser Science, 22761 Hamburg, Germany}

\author{Martin Claassen}
\altaffiliation{claassen@sas.upenn.edu}
\affiliation{Department of Physics and Astronomy, University of Pennsylvania, Philadelphia, PA 19104}

\author{Angel Rubio}
\affiliation{Max Planck Institute for the Structure and Dynamics of Matter, Center for Free Electron Laser Science, 22761 Hamburg, Germany}
\affiliation{Center for Computational Quantum Physics, Simons Foundation Flatiron Institute, New York, NY 10010 USA}

\author{Dante M. Kennes}
\altaffiliation{dante.kennes@rwth-aachen.de}
\affiliation{Institut f\"ur Theorie der Statistischen Physik, RWTH Aachen University and JARA-Fundamentals of Future Information Technology, 52056 Aachen, Germany}

\affiliation{Max Planck Institute for the Structure and Dynamics of Matter, Center for Free Electron Laser Science, 22761 Hamburg, Germany}

\date{\today}

\begin{abstract}
We study the electronic structure and correlated phases of twisted bilayers of platinum diselenide using large-scale ab initio simulations combined with the functional renormalization group. PtSe$_2$ is a group-X transition metal dichalcogenide, which hosts emergent flat bands at small twist angles in the twisted bilayer. Remarkably, we find that moir\'e engineering can be used to tune the strength of Rashba spin-orbit interactions, altering the electronic behavior in a novel manner. We reveal that an effective triangular lattice with a twist-controlled ratio between kinetic and spin-orbit coupling scales can be realized. Even dominant spin-orbit coupling can be accessed in this way and we discuss consequences for the interaction driven phase diagram, which features pronounced exotic superconducting and entangled spin-charge density waves. 
\end{abstract}
\maketitle

\section{Introduction}
The advent of moir\'e heterostructures and the demonstration of superconducting, correlated insulating and topological phases of matter in these materials~\cite{cao2018unconventional,cao2018mott,lu2019superconductors,Cao2020strange,Polshyn2019,yankowitz2019tuning,liu2021tuning,stepanov2020untying, arora2020superconductivity, Zondiner2020,Wong2020,Xie2019spectrosopic,Kerelsky2019maximized,Jiang2019charge,Choi2019correlations,cao2020nematicity}, has triggered a surge of  theoretical and experimental studies. Common to these is the idea that a slight lattice constant mismatch or rotation between adjacent layers of two-dimensional van der Waals materials can significantly quench kinetic energy scales and alter the effective electronic band structure that dictates the low-energy behavior. Moir\'e heterostructures were envisioned to allow novel control of the ratio between kinetic and potential energies by superlattice engineering, allowing an exploration of strong electronic correlations in a tunable condensed matter setting \cite{Kennes2021}.
Recent experimental findings suggest that superconductivity in twisted sheets of bilayer graphene is indeed of unconventional nature~\cite{oh2021evidence}
and that twisted trilayer graphene may favor triplet pairing~\cite{Park2021,cao2021large,hao2021electric,kim2021spectroscopic}.
Substantial efforts have been made to unravel the nature of correlated states in related graphitic moir\'e materials. Experiments report, among others, correlated states in twisted mono-bilayer graphene~\cite{chen2020electrically,shi2020tunable}, twisted double bilayer graphene~\cite{liu2020tunable,shen2020correlated,cao2019electric,tutuc2019,RubioVerdu20} and rhombohedral graphene aligned with hexagonal boron-nitride~\cite{Chen2019ABC,chen2019evidence,chen2020tunable}. A vast amount of theoretical work predicts correlation effects for an even larger subspace of the possible twisted graphitic moir\'e systems~\cite{fischer2021unconventional,fischer2021spin,Lee2019,schrade2021nematic,xian2021engineering,khalaf2021charged,kleblABC,cea2021superconductivity,soriano2020exchange,wolf2019electrically,liu2021,qin2021,kezilebieke2020moireenabled,gonzalez2019kohn,you2019superconductivity,khalaf2019magic}.
In addition, in twisted bilayer graphene, control over topological properties has already been demonstrated \cite{nuckolls2020strongly,xie2021fractional,pierce2021unconventional,stepanov2020competing,choi2020tracing,sharpe2019emergent,wu2021chern,Das2020Chern,park2020flavour, saito2020independent} showing correlated Chern insulating phases.  

\begin{figure*}[t]
	\centering
	\includegraphics[width=\textwidth]{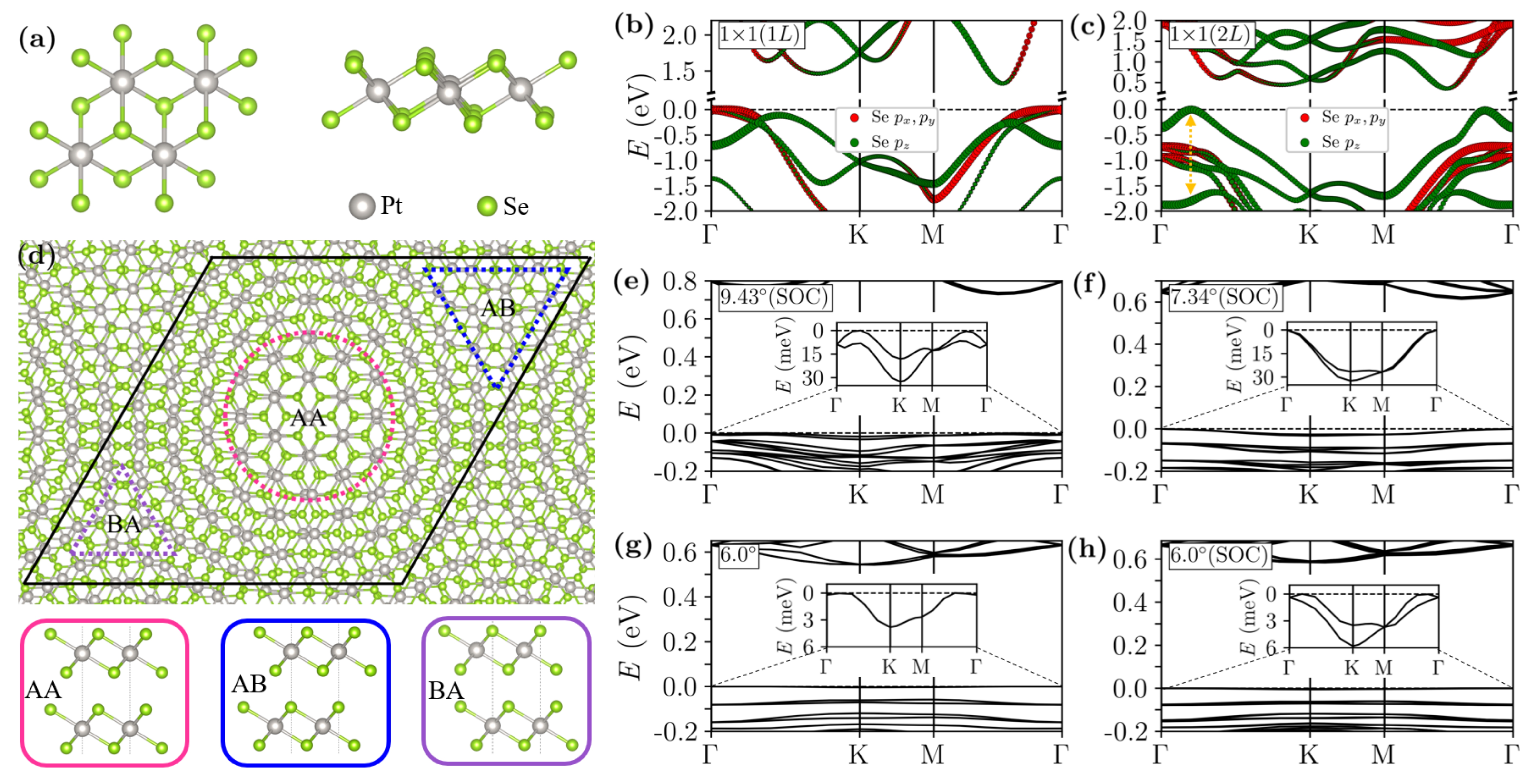}
	\caption{\captiontitle{Atomic and electronic structure of PtSe$_2$ layers.} {\bf (a)} Top (left) and side (right) views of monolayer PtSe$_2$ in the 1T type structure. {\bf (b,c)} Band structures of monolayer {\bf(b)} and bilayer {\bf(c)} PtSe$_2$ in its 1x1 primitive cell. The top of the valence bands are dominated by Se $p_z$ orbitals. Compared with the monolayer, interlayer coupling induces a huge splitting of these $p_z$ orbitals in the bilayer (indicated by the arrow). {\bf (d)} Atomic structure of twisted bilayer PtSe$_2$  \append{at a twist angle of $6^\circ$} with various local stacking of AA, AB and BA. \append{{\bf (e-h)} Band structures of twisted bilayer PtSe$_2$ at 9.43$^\circ$ {\bf(e)}, 7.34$^\circ$ {\bf(f)} and 6$^\circ$ {\bf(h)} with SOC and 6$^\circ$ {\bf(g)} without SOC.}}
	\label{fig:dft}
\end{figure*}

Beyond graphene, experiments have studied twisted sheets of transition metal dichalcogenides (TMDs), concentrating primarily on group-VI homo- or hetero-bilayers of MoS$_2$/Se$_2$ or WeS$_2$/Se$_2$, with fascinating observations of strongly correlated behavior~\cite{wang2020correlated,scherer2021,kiese2021tmds,tang2020,xian20, regan2020,witt2021doping,Vitale2021,zang2021,malic2018dark,palummo2015exciton} and
excitonic physics~\cite{nayak2017probing,wang2018colloquium,Rivera2018,alexeev2019resonantly,Andersen19,Jin18,Brem2020}. Further proposals for TMD materials include exotic superconducting states with topological features~\cite{schrade2021nematic,scherer2021}, possible spin-liquid phases~\cite{kiese2021tmds} and engineering of multi-orbital systems in group-IV TMDs as a realization of the elusive Kagome lattice with strong and tunable spin-orbit coupling (SOC) which exhibits fractional quantum anomalous Hall and Chern insulating states~\cite{claassen2021zrs2}. 

However, going beyond graphene- and TMD-based systems, the profusion of available van der Waals materials allows for even more exotic quantum materials design. For instance, the reduced rotation symmetry of monochalcogenides permits engineering quasi-one dimensional structures~\cite{kennes19}; alternatively, by departing from the realm of few-layer systems, moir\'e induced control of three dimensional materials becomes possible~\cite{xian2021engineering}. By considering oxides as a basis for moir\'e engineering, exotic $d$-wave superconductivity is supposed to emerge~\cite{Can2021}  with a potential connection to the fascinating high-$T_\mathrm{c}$ phase of the cuprates.  All of this is to show that in the field of moir\'e engineering much more is expected to be possible by exploiting the different chemical compositions offered by the choice of materials to consider. This general concept of identifying novel phenomena to be controlled by moir\'e engineering might culminate into a versatile new solid state-based platform~\cite{Kennes2021} to access quantum materials behavior with unprecedented level of tunability. The discovery and characterization of limits and opportunities in new van-der-Waals materials platforms hence remains an important avenue of pursuit.

Here, we add to the catalog of phenomena realizable by moir\'e engineering by considering the group-X TMD PtSe$_2$, which is exfoliable down to monolayer~\cite{bae2021exciton}. It has raised lots of interest for its outstanding optical and electrical properties and high air-stability~\cite{zhao2017high,wang2021layered}. In the context of  moir\'e engineering, this material is interesting due to the substantial spin-orbit coupling of heavy transition metal ions. We demonstrate via an ab-initio characterization of large unit-cell systems at small twist angles that, when the kinetic energy scales are quenched by twisting two sheets of PtSe$_2$ with respect to each other, a controlled twist-dependent tuning of Rashba SOC and kinetic energy scales can be achieved. Surprisingly, we find that relatively large twist angles of about 6$^\circ$ are sufficient to quench kinetic energy scales small enough to promote SOC to be the dominant energy scale. In contrast to the strong SOC of twisted bilayers of ZrS$_2$~\cite{claassen2021zrs2}, the SOC interaction in PtSe$_2$ is mainly of Rashba type and relies existentially on broken inversion symmetry in the moir\'e superstructure, hence realizing a new regime. We discuss consequences for correlated phases of matter using a weak-coupling functional renormalization group approach, which can be viewed as an unbiased renormalization-group-enhanced random phase approximation. Our results indicate a rich phase diagram of intertwined charge-spin density waves, which in the case of SOC cannot be disentangled, and exotic mixed-parity superconducting phases with topologically non-trivial properties.

The paper is structured as follows: We start from a full ab-initio characterization of the twist angle dependence of the electronic band structure including the spin-orbit coupling for twisted bilayers of PtSe$_2$. We demonstrate the twist-dependent reduction of the effective electronic bandwidth, which coincides with the emergence of strong Rashba interactions. The resulting moir\'e bands span a triangular lattice with few nearest neighbor hoppings plus Rashba SOC. We then treat this model by adding a Hubbard interaction and outline the emerging phase diagram. A discussion concludes the paper, with details of the Methods used appended below.

\begin{figure*}
	\centering
	\includegraphics[width=\textwidth]{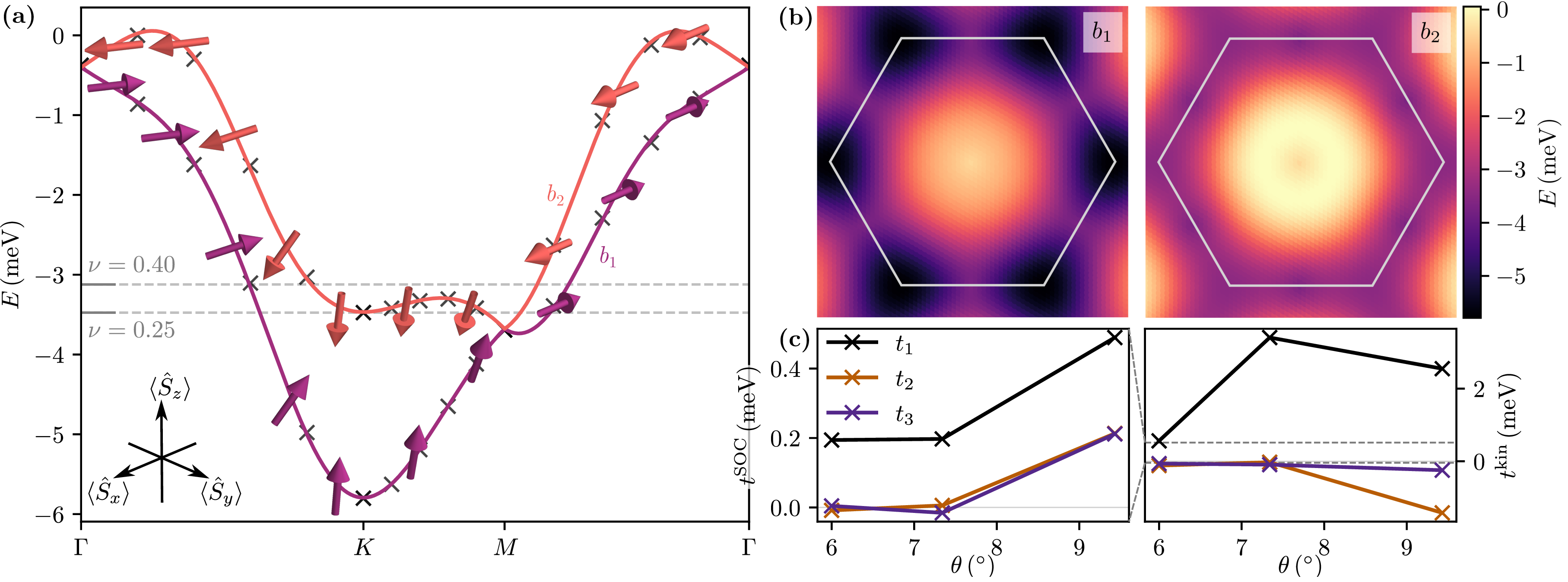}
	\caption{\captiontitle{Tight-binding model for twisted PtSe$_2$ at $\theta=6^\circ$. (a)} Tight binding band structure (line) fitted to DFT band structure (black crosses). The small arrows indicate the spin expectation value as a function of lower band $b_1$, upper band $b_2$ and momentum. We indicate the flat band filling values $\nu=0.25$ and $\nu=0.40$ as dashed horizontal lines. {\bf(b)} Dispersion of the tight-binding model in the full BZ for both bands. The BZ boundaries are indicated with the light gray hexagons. {\bf(c)} Hopping parameters obtained from fit to DFT results as a function of twist angle $\theta$. For $\theta=6^\circ$, the nearest neighbor Rashba SOC  magnitude is $\sim40\%$ of the nearest neighbor non-SOC hopping indicating stark SOC enhancement.}
	\label{fig:tb}
\end{figure*}

\section{Ab-Initio Characterization}
To provide a first-principle characterization of the electronic structure of twisted bilayers of PtSe$_2$ we first employ a density functional theory (DFT) based approach to the  material, which at small twist angles can exhibit a very large unit cell (see Methods). As Pt is a heavy element, spin-orbit coupling is important and it is included in the calculations, which breaks the underlying $SU(2)$ spin rotation symmetry (see Methods for more details). PtSe$_2$ is a group-X transition metal dichalcogenide, and the atomic structure of a monolayer is visualized in Fig.~\ref{fig:dft} {\bf (a)} as a top and side view. The Pt and the Se atoms are shown as grey and green spheres, respectively. We concentrate on the energetically stable 1T lattice structure of PtSe$_2$. In panels {\bf (b)} and {\bf (c)} we report the DFT bands structure of the monolayer {\bf(b)} and the untwisted bilayer {\bf (c)} for reference. As shown in the figures, the states at the valence band edge are dominated by Se $p_x$, $p_y$ and $p_z$ orbitals. Among them, we find that the Se $p_z$ states are very sensitive to interlayer coupling, which strongly hybridize in the bilayer and form bonding and anti-bonding states with a large energy splitting (indicated by the orange arrow in Fig.~\ref{fig:dft} {\bf (c)}). Consequently, the top of the valence bands shift from the Se $p_x$ and $p_y$ states in the monlayer to the Se $p_z$ states in the bilayer. This is different from the case of another 1T TMD, ZrS$_2$ \cite{claassen2021zrs2}, in which the top of the valence bands are dominated by $p_x$ and $p_y$ states, in both monolayer and bilayer. Because of the strong interlayer coupling of the Se $p_z$ states at the valence band edge, we expect that flat bands can be formed in twisted bilayer of PtSe$_2$ at relatively large twist angles.

Next we turn to non-zero twist angles. Panel {\bf (d)} shows the real space moir\'e pattern emerging for twisted PtSe$_2$. We mark different regions as `AA', `AB' and `BA' stacking, with the local stacking arrangement of the atoms given as insets to the side. In panels {\bf (g)} and {\bf (h)} we report the DFT analysis of the band structure in the twisted system excluding {\bf (g)} and including {\bf (h)} SOC \append{at a twist angle of $6^\circ$. Panels~{\bf(e,f)} show the DFT band structure including SOC at twist angles $9.43^\circ$ and $7.34^\circ$, respectively.} As twist angles $\sim 6^\circ$ are approached, the electronic band near the Fermi energy becomes very flat with a width of $\sim 4$ (6) meV in the calculations without (with) SOC. This is significantly lower than the corresponding bandwidth for twisted bilayer graphene \cite{cao2018a}, especially at such relatively large twist angle.  Comparing the calculations with and without SOC, the relevance of including the latter becomes strikingly clear. The degeneracy of the up- and down-spin electronic bands without SOC is lifted by including the SOC in the calculations, by virtue of broken inversion symmetry. To quantify this effect we will next analyze the relevance of SOC versus kinetic energy scales in dependence of the twist angle using a tight-binding \substitute{ansatz}{approach}. \append{As the bandwidth of twisted bilayer PtSe$_2$ at $6^\circ$ is small enough for correlation effects to be relevant, we do not perform further DFT calculations for smaller twist angles, which are increasingly expensive. Nevertheless, we expect the bandwidth could be further reduced at smaller angles.}

\section{Tight-Binding Description and Strong Spin-Orbit Coupling}
We model the electronic flat bands with a tight-binding model on the triangular lattice, taking into account hopping parameters connecting up to third nearest neighbors. Due to the combination of broken inversion and $SU(2)$ spin rotation symmetry, we capture strong spin orbit coupling of the flat bands using a Rashba term. Additionally, the non-$SU(2)$ nature of the system can lead to intrinsic, spin-dependent electric field effects described by a complex spin-dependent phase accompanying the kinetic hopping parameters \cite{kane-mele2005topological,andy2021hartree}. The kinetic part of the Hamiltonian then reads
\begin{multline}
    H^\mathrm{kin} = \sum_{\sigma} \sum_{B_{ij}} \sum_{\bvec b_{ij}\in B{ij}} \big( t^\mathrm{kin}_{ij}e^{i\phi_{ij}\sigma}c^\dagger_{i\sigma}c^{\phantom\dagger}_{j\sigma}
+ {\rm h.c.}   \big)\,,
\label{eq:kanemele}
\end{multline}
where $B_{ij}$ is a set of $C_3^z$ symmetry related directed bonds $\bvec b_{ij}$ in the triangular lattice with equal length, $t^\mathrm{kin}_{ij}$ are the kinetic hopping parameters and $\phi_{ij}$ the Kane-Mele phase factors. By construction, this Hamiltonian fulfills both time reversal and $C_3^z$ symmetry. The Rashba term is given by
\begin{equation}
    H^\mathrm{SOC} = i\,\sum_{ij\sigma\sigma'}t^\mathrm{SOC}_{ij}\big(\hat{\bvec\sigma}\times\bvec b_{ij}\big)^{\sigma\sigma'}_z\,c^\dagger_{i\sigma}c^{\phantom\dagger}_{j\sigma'}\,,
\label{eq:rashba}
\end{equation}
with $t^\mathrm{SOC}_{ij}$ the SOC hopping parameters. Finally, we include a chemical potential ($\mu$), such that the full tight binding Hamiltonian becomes
\begin{equation}
    H^0 = H^\mathrm{kin} + H^\mathrm{soc} - \mu\,\sum_{i\sigma} c^\dagger_{i\sigma}c^{\phantom\dagger}_{i\sigma}\,.
    \label{eq:tb}
\end{equation}
\append{Note that we do not account for changes in tight-binding parameters when varying the chemical potential, such that a change in $\mu$ is directly reflected in a change in filling $\nu$.}

Figure~\ref{fig:tb}~{\bf(a)} shows both the DFT band structure (black crosses) and the tight-binding fit (line) for a twist angle of $\theta=6^\circ$. We list the fit parameters for  $6^\circ$ as well as for two additional twist angles ($7.34^\circ$ and $9.43^\circ$) that we calculated using DFT \append{[cf.~Fig.~\ref{fig:dft}~{\bf(e,f)}]},  in Tab.~\ref{tab:hopparams} in the Methods section.
Additionally, we show the spin expectation value as three-dimensional arrows in Fig.~\ref{fig:tb}~{\bf(a)}: Along the path $\Gamma$--$K$, the spin has finite expectation value in the $x$-$y$ plane which then gradually shifts towards spin-$z$ at the $K$ point. From $K$ to $M$, the expectation value of the spin-$z$ component is nonzero with a slight tilt towards the $x$-$y$ plane close to $M$. From $M$ to $\Gamma$, the expectation value fully lies in spin-$x$ direction. The possibility to generate a finite expectation value of a specific spin component aside from $S_z$ arises from making a specific choice of the part $\Gamma$--$M$ of the irreducible path. We label the lower band by $b_1$ and the upper band by $b_2$ with Fig.~\ref{fig:tb}~{\bf(b)} showing a two-dimensional false color plot of the dispersion for the same twist angle. The hexagonal BZ is indicated as gray lines. Strong spin orbit coupling and the very small band width of approximately $6\,\mathrm{meV}$ are clearly visible. The only degeneracy points of the band structure lie on the BZ boundary at the $M$ points and at $\Gamma$.

After having established an accurate tight-binding representation of our DFT results, we can quantify the strength of SOC as a function of twist angle by carrying out the fitting procedure at two other (commensurate) twist angles. The resulting kinetic and SOC hopping parameters $t_1$, $t_2$, $t_3$ are shown in Fig.~\ref{fig:tb}~{\bf(c)}. 
For all three twist angles considered in the scope of this work, we see that SOC is extremely relevant. Since the moir\'e potential becomes increasingly relevant at smaller twist angles, the overall kinetic energy scale given by the nearest neighbor hopping $t_1^\mathrm{kin}$ is drastically reduced for $\theta=6^\circ$. As a consequence, the SOC hopping parameter $t_1^\mathrm{SOC}$ is around $40\%$ of the non-SOC $t_1$. Furthermore, the influence of longer range hoppings $t_2$ and $t_3$ becomes smaller when decreasing the twist angles.

With this in mind, we continue our analysis of the tight binding model and complement the quenched kinetic energy Hamiltonian $H^0$ with onsite Coulomb interactions $H^U$:
\begin{equation}
    H^U = U\,\sum_{i\sigma\delete{\sigma'}} c^\dagger_{i\sigma} c^\dagger_{i\substitute{\sigma'}{\bar{\sigma}}} c^{\phantom\dagger}_{i\substitute{\sigma'}{\bar{\sigma}}} c^{\phantom\dagger}_{i\sigma}\,.
\end{equation}
In the following, we will study the effect of $H^U$ on the non-interacting moir\'e Hamiltonian $H^0$. \append{We therefore focus on $\theta=6^\circ$ as twist angle for two reasons. First, kinetic energy scales are strongly suppressed and second, the quality of the tight-binding fit is the best due to long range hoppings being least relevant (among the cases studied within this work).}

\begin{figure*}[t]
	\centering
	\includegraphics[width=\textwidth]{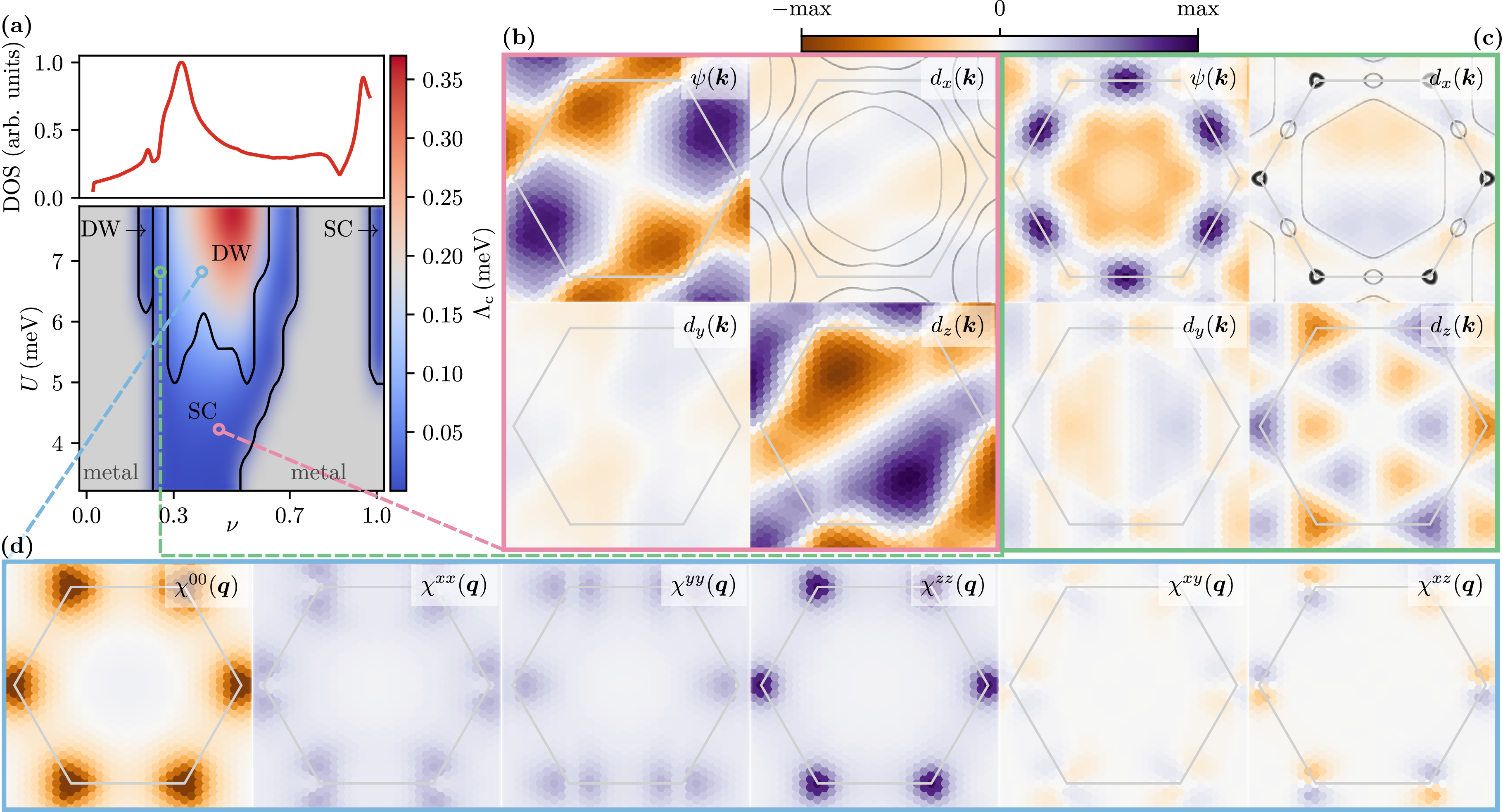}
	\caption{\captiontitle{Functional renormalization group analysis of $6^\circ$ twisted PtSe$_2$. (a)} Phase diagram calculated on a regular $21\times8$ grid in filling $\nu$ and moir\'e-Hubbard interaction parameter $U$. Upper panel: density of states as a function of $\nu$. The central van-Hove peak drives the density wave (DW) and superconducting (SC) regions around $\nu=0.4$ (lower panel). The second van-Hove singularity at $\nu\approx1.0$ leads to a much smaller region of superconductivity. The color encodes the critical scale $\Lambda_\mathrm{c}$ that roughly corresponds to a transition temperature. Regions where the fRG does not predict ordering are labeled by `metal'. {\bf(b)} Superconducting gap obtained from solving a linearized gap equation with the fRG vertex $\Gamma^{(4)}$ at the critical scale $\Lambda_\mathrm{c}$ for $U=5.27\,\mathrm{meV}$ and $\nu=0.4$. The superconducting gap $\Delta_{\sigma\sigma'}(\bvec k)$ is transformed to singlet [$\psi(\bvec k)$] and triplet [$\bvec d(\bvec k)$] space and can be chosen to be real in this basis. We show only one of the two degenerate instabilities with intertwined $p$ and $d$ wave symmetries in the respective coupled singlet and triplet components. The BZ is indicated with the gray hexagon. {\bf(c)} Analogous analysis for the second SC instability found at filling $\nu=0.25$. This solution is not doubly degenerate and has $g$-wave and $f$-wave symmetry in the respective singlet and triplet channels. For both superconducting instabilities {\bf(b)} and {\bf(c)}, most of the triplet weight is concentrated in $d_z(\bvec k)$. The respective Fermi surfaces are indicated in the top right subpanel as black lines. {\bf(d)} DW instability belonging to the central patch at $U=7.61\,\mathrm{meV}$ and $\nu=0.6$. The density channel $\chi^{00}(\bvec q)$ is strongly coupled with the spin-$z$ channel $\chi^{zz}(\bvec q)$ as a result of the material's substantial SOC. For this filling, the transfer momentum is commensurate at $\bvec q=K$ and $\bvec q=K'$. Within the DW region that $\nu=0.6$ belongs to [cf.~{\bf(a)}], the ordering vector becomes slightly incommensurate around $K$ and $K'$ for other values of $\nu$.}
	\label{fig:frg}
\end{figure*}

\section{interaction-Driven Phases of Matter}
We approach the interacting quantum many-electron problem  using the unbiased functional renormalization group (fRG) \cite{Metzner2012a}. The broken $SU(2)$ symmetry renders even this two-band problem a significant challenge and we truncate the infinite hierarchy of flow equations set up within the fRG approach at the four-point vertex $\Gamma^{(4)}$. Furthermore, focusing on static quantities, we neglect frequency dependencies of $\Gamma^{(4)}$ and further set the two-point vertex (self-energy) to zero. The fRG flow then amounts to solving a differential equation (see Methods) for $\Gamma^{(4)}$ as a function of $\Lambda$, the parameter that smoothly interpolates from the free theory at $\Lambda=\infty$ to the full, interacting theory at $\Lambda=0$. During the flow, we search for divergences in $\Gamma^{(4)}$ indicating a tendency towards long-range order. The four-point vertex is then analyzed at the final scale $\Lambda_\mathrm{c}$ that roughly corresponds to a critical temperature of the phase transition associated with the divergence.

Within our approach we can distinguish between charge/spin-density wave (DW) or superconducting (SC) instabilities. The primary indicator for the type of divergence is given by the divergent channel during the fRG flow, which can either be of particle-particle (SC) or particle-hole (DW) type. If the vertex remains finite up to $\Lambda=0$, the fRG does not predict long range order and thus a metallic phase.

Figure~\ref{fig:frg}~{\bf(a)} shows the resulting phase diagram for the flat bands of twisted bilayer PtSe$_2$ at $\theta=6^\circ$ as a function of moir\'e-Hubbard interaction strength $U$ and filling factor $\nu$ \append{parametrized by the chemical potential $\mu$}. The upper panel displays the system's density of states (DOS) with two main van Hove singularities at $\nu\approx0.4$ and $\nu\approx1.0$. These regions of high DOS are responsible for the instabilities. However, the lower panel reveals that both SC and DW ordering can emerge away from points with divergent DOS. Moreover, there is a rich phase structure with various regions of SC and DW order at a broad range of critical scales (encoded in color). The DW instabilities predominantly occur at large $U$, whereas the SC instabilities are \emph{dominant} for a broad range of fillings ($\nu=0.25 \sim 0.5$) and interactions ($U=3.5 \sim 5\,\mathrm{meV}$), and are only flanked by metallic regions. At larger interaction strengths, we observe a second DW instability with very low critical scale at $\nu=0.2$ and a second superconducting instability driven by the van Hove peak at $\nu=1.0$.

\begin{figure*}
    \centering
    \includegraphics[width=\textwidth]{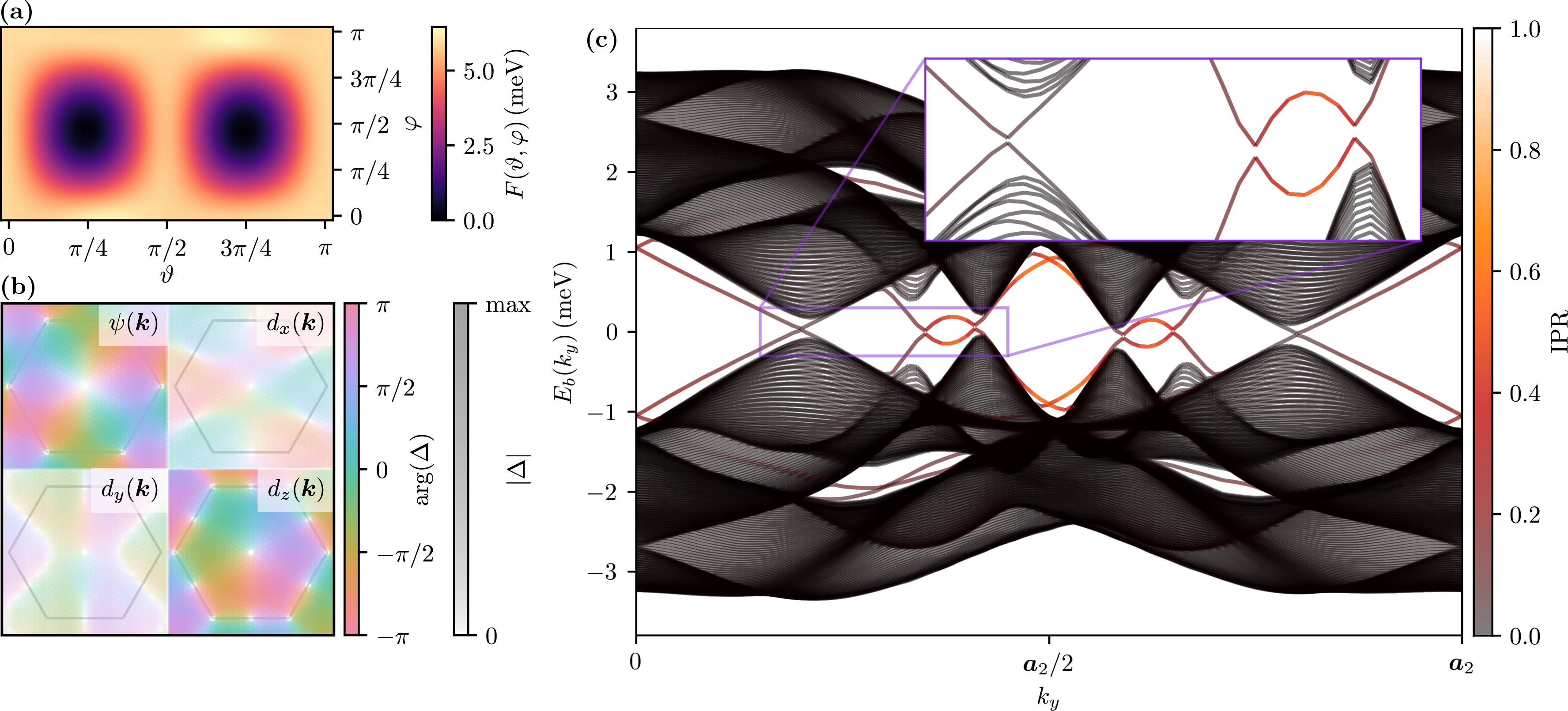}
    \caption{\captiontitle{Topological superconductivity in $6^\circ$ twisted PtSe$_2$. (a)} Minimization of free energy as a function of superposition angles $\vartheta$ and $\varphi$ for $p/d$-wave gap at $\nu=0.4$. The free energy is minimized at $\vartheta_0\approx\pi/4$ and $\varphi_0\approx\pi/2$. {\bf(b)}~Chiral gap function in singlet and triplet space for the linear combination that minimizes the free energy. {\bf(c)}~Half real space band structure with open boundary conditions in $\bvec a_1$ direction (cylinder geometry) at $\nu=0.4$ and SC gap from {\bf(b)}. The line color encodes the inverse participation ratio, with a clear signature of localization for the modes in the superconducting gap.}
    \label{fig:topo}
\end{figure*}

Figures~\ref{fig:frg}~{\bf(b,c)} illustrate the two types of superconducting instabilities found. First, for most of the central SC region and the remote region at $\nu=1.0$ in the phase diagram [Fig.~\ref{fig:frg}~{\bf(a)}], the leading instability is of $d$-wave and $p$-wave type [see Fig.~\ref{fig:frg}~{\bf(b)}]. The $d$ and $p$ wave symmetries mandate that the superconducting gap is doubly degenerate (second instability not shown). We plot the superconducting gap amplitude in the BZ for both the singlet [$\psi(\bvec k)$] and triplet [$\bvec d(\bvec k)$] channel. In the case of strong SOC and lack of inversion symmetry, the decoupling of a superconducting order parameter into independent singlet and triplet components is impossible and mixed-parity superconducting states form. Nevertheless, we can transform the superconducting gap $\Delta_{\sigma\sigma'}(\bvec k)$ to singlet and triplet space (see Methods) \cite{sigrist1991phenomenological,smidman2017superconductivity}, but with instabilities that have weight in both spaces at the same time. In momentum space, $\psi(\bvec k)$ and $\bvec d(\bvec k)$ must fulfill the (anti-)symmetry relations of singlet (triplet) gaps. Second, for the SC instabilities at $\nu=0.25$, we find a different order parameter [see Fig.~\ref{fig:frg}~{\bf(c)}] with dominant $g$-wave (singlet) and $f$-wave (triplet, $d_z$) components and little weight in the $d_x$ and $d_y$ components of the triplet vector.
This order parameter is not degenerate and, by its $g/f$-wave symmetry, leads to a nodal superconducting state.

To gain an intuitive understanding of the weight distribution in the $\bvec d$-vector, it is helpful to consider the spin polarization of the Fermi contour. In the presence of strong spin-orbit coupling [Eq.~\eqref{eq:rashba}] and time-reversal symmetry ($\mathcal{T}$) an arbitrary single particle state $\ket{\bvec k_F, \uparrow}$ at Fermi momentum $\bvec k_F$ and with spin $\sigma = \uparrow$ measured relative to the spin-polarization axis at $\bvec k_F$ can be transformed to $\mathcal{T} \ket{\bvec k_F, \uparrow} = \ket{-\bvec k_F, \downarrow}$. Since inversion symmetry $\mathcal{I}$ is no longer conserved in the system, the states $\ket{\bvec k_F, \uparrow}$ and $\ket{-\bvec k_F, \uparrow}$ are no longer required to be degenerate~\cite{smidman2017superconductivity} such that opposite spin Cooper pairs are favored in this situation~\cite{millis2020}. Indeed, we observe that at filling $\nu=0.4$, the bands at the Fermi energy are mostly $z$-polarized
leading to dominant weight in the singlet $\psi(\bvec k)$ and triplet $d_z (\bvec k)$ component, while the other components of the $\bvec d$-vector are substantially suppressed. As the system is filled with more electrons, the spin polarization axis changes from $z$ to the $x$-$y$ plane and consequently shifts the weight in the $\bvec d$-vector.

At and close to the central van Hove singularity, the system is susceptible to an intertwined magnetic/charge density order (i.e. divergence in the particle-hole channel) presented in Fig.~\ref{fig:frg}~{\bf(d)}. The physical spin- and density channels $\chi^{lm}(\bvec q)$ with $l,m\in\{0,x,y,z\}$ are obtained from the four-point particle-hole susceptibility $\chi^\mathrm{PH}_{\sigma_1,\sigma_2,\sigma_3,\sigma_4}(\bvec q)$ which is in turn calculated from the vertex $\Gamma^{(4)}$ at the critical scale $\Lambda_c$ (see Methods) \cite{scherer2018spin-orbit}. By virtue of the strong SOC, the density-density response $\chi^{00}(\bvec q)$ (first panel) and the spin-spin responses are intrinsically coupled. Most weight in the spin sector is in $\chi^{zz}(\bvec q)$, i.e. the spin-$z$ response with dominant ordering vectors $K$ and $K'$. The in-plane responses are much weaker; albeit they are non-negligible. $\chi^{yz}(\bvec q)$ is not shown in Fig.~\ref{fig:frg}~{\bf(d)}, though its form can be inferred from $\chi^{xz}(\bvec q)$ by symmetry. The less dominant DW instability at $\nu=0.2$ is of qualitatively different type with dominant terms in the density-density and $\chi^{zz}$ sectors for transfer momentum $\bvec q=0$.

We further investigate the physical consequences that arise when the system is in the intertwined $d/p$-wave superconducting phase at $\nu=0.4$. To examine which linear combination of the two degenerate instabilities is energetically favored, we calculate the free energy in the superconducting phase (see Methods) as a function of all possible complex superpositions of the two $d/p$-wave order parameters:
\begin{equation}
    \hat{\Delta}^{\vartheta,\varphi}(\bvec k) = \cos\vartheta e^{i\varphi}\hat{\Delta}^1(\bvec k) + \sin\vartheta\hat{\Delta}^2(\bvec k) \,.
    \label{eq:frg-complex-sc}
\end{equation}
Figure~\ref{fig:topo}~{\bf(a)} indicates that the free energy in the superconducting phase is minimized for $\vartheta_0\approx\pi/4$ and $\varphi_0\approx\pi/2$.
This particular choice of $\vartheta_0$ and $\varphi_0$ leads to an order parameter that preserves $C_3$ rotational symmetry in the $\psi(\bvec k)$ and $d_z(\bvec k)$ components while breaking time reversal symmetry.
We display the associated chiral superconducting order parameter in Fig.~\ref{fig:topo}~{\bf(b)}, where we encode the complex phase as color and the magnitude as lightness. Next, we set an amplitude of $|\Delta|_\mathrm{max}=2.35\,\mathrm{meV}$ and study the Bogoljubov-de-Gennes (BdG) bandstructure $E_b(\bvec k)$ in Fig.~\ref{fig:topo}~{\bf(c)}. To assess whether the system is topologically nontrivial, we first numerically diagonalize the BdG Hamiltonian (see Methods) on a cylindrical geometry. We periodically continue the system in $\bvec a_2$ direction and open the boundary in the $\bvec a_1$ direction. Further, we color-code the inverse participation ratio (IPR) as a function of band index $b$ and momentum in $\bvec a_2$ direction ($k_y$). Additionally, we determine the Chern number of the two upper BdG bands, where $C=+2$ and the two lower BdG bands with $C=-2$. This leads us to conclude that the superconducting order is topologically non-trivial.

\section{Discussion}
Our results elevate twisted bilayer PtSe$_2$ as a novel platform for engineering strong Rashba spin-orbit coupling in a tunable setting. Importantly, the strong spin-orbit coupling regime can be accessed in a controllable fashion, allowing a novel inroad into this evasive physical regime.  We discussed consequences of the strong spin-orbit coupling as well as the exotic form of the engineered low-energy effective Hamiltonian, which shows prominent effects of the breaking of the $SU(2)$ symmetry even without interactions. The bands found within our approach have non-trivial spin polarization and an intriguing spin-momentum locking of potential interest for novel nano-devices and spintronics \cite{Spintronics1,Spintronics2}. 

The physics becomes even more rich upon the inclusion of electronic interactions.
\append{We focus the analysis on the twist angle $\theta=6^\circ$ motivated by the fact that the tight binding fit dictating the low-energy band structure has highest quality in this case because long range hopping parameters become less relevant. Furthermore, the flat-band bandwidth for $\theta=6^\circ$ is the smallest among the twist angles considered in this work which leads to substantially quenched kinetic energy scales and enhanced interaction effects.}
The system exhibits two separate van Hove singularities which trigger a series of unconventional weak-coupling instabilities. We scrutinized these instabilities using unbiased renormalization group enhanced diagrammatic techniques, which point to extended regions where density waves or superconductivity emerge.  Since spin and charge are entangled in non-$SU(2)$ symmetric systems without inversion symmetry, the phase diagram and the classification of the phases of matter expected becomes extremely intricate. Our analysis shows that of the two types of superconductivity present in the phase diagram, one is topologically non-trivial and the other is trivial. The topologically trivial superconducting phase, which occurs only at low filling fractions of the flat bands, is still interesting for its high-angular-momentum form factors, being of the $g$- and $f$-type.  The topologically non-trivial superconducting phase occupies a large fraction of the phase diagram as a function of filling (around half-filling) and interaction strength, which suggests PtSe$_2$ as an interesting material to search for topological superconductivity.

Our work highlights another exciting example enabled by flexible moir\'e engineering, concentrating this time on the less explored tailoring of spin-orbit coupling. Engineering spin-orbit coupling is an important topic in the field of quantum materials as spin-orbit coupling can trigger many fascinating topological transitions which might find a materials based application in moir\'e materials for the first time.

\section{Methods}
{\it Density functional theory} --- In our characterization of the large unit-cell twisted bilayer PtSe$_2$ material we used the Vienna Ab initio simulation package (VASP). VASP was employed to determine the ground state of the system within the density functional theory (DFT) \cite{kresse93ab} with the basis chosen to be plane waves and energy cutoff of 400 eV. The pseudo potentials are generated using the projector augmented wave method (PAW) \cite{blochl94} and the exchange-correlation functions are treated within the Perdew, Burke, and Ernzerhof (PBE) \cite{pbe}. We calculate the equilibrium lattice parameters of PtSe$_2$ in the bulk phase and found that the optB86b van der Waals (vdW) functionals \cite{klimevs2009chemical} provide better agreement with the experimental values, within less than 2$\%$ errors \cite{kliche1985far}. The optB86b vdW functionals are then adopted for all calculations. 
For these very large uni-cell simulations a 1x1x1 momentum grid is used to characterize the ground state and the mechanical relaxation. We construct the supercell of the considered bilayer system by using the optimized lattice constants of a 1x1 unit cell. 
DFT is most conveniently set up using periodic boundary condition and therefore, \substitute{a long}{along} the z-direction an auxiliary vacuum region larger than $15\,\text{\AA}$ is added. This region is chosen large enough that artificial interaction between the periodic slabs can be neglected. Our calculations are fully relaxed (w.r.t. all the atoms), which is known to be important in other moir\'e systems to avoid artificial effects stemming from unrelaxed structures \cite{walet2019,lucignano2019,jain2016}. The relaxation procedure ensures that the force on each atom converges to values smaller than $0.01\,\text{eV/\AA}$.
For all calculations, due to the relativistic effect in heavy element Pt,  the spin-orbit coupling (SOC) effect is considered while results without SOC are also calculated for comparison to estimate the effect of the SOC on the moir\'e flat bands.
The twisting angles of 6$^\circ$, 7.34$^\circ$ and 9.43$^\circ$ contain in total 546, 366 and 222 number of atoms to consider in the unit cell. 

{\it Tight-binding parameters} --- We perform fitting of the tight-binding Hamiltonian [Eq.~\eqref{eq:tb}] to DFT band structures of the moir\'e flat bands of twisted PtSe$_2$. By that, we obtain the ten parameters $\mu, t^\mathrm{kin}_{1,2,3}, t^\mathrm{SOC}_{1,2,3}$ and $\phi_{1,2,3}$ for each of the three twist angles considered. We tabulate the parameters in Tab.~\ref{tab:hopparams}.
\begin{table}[ht]
    \centering
    \caption{\captiontitle{Tight-binding fit parameters of the moir\'e flat bands of twisted PtSe$_2$ for three twist angles.}}
    \label{tab:hopparams}
  \begin{tabular}{|c||c|c|c|c|}
    \hline\hline
    $\theta$ & $\mu\append{\,(\mathrm{eV})}$ & $t_1^\mathrm{kin}\,(\mathrm{\delete{m}eV})$ & $t_2^\mathrm{kin}\,(\mathrm{\delete{m}eV})$ & $t_3^\mathrm{kin}\,(\mathrm{\delete{m}eV})$ \\
    \hline\hline
    $6.00^\circ$ &
        $2.52\cdot10^{-3}$ & $5.62\cdot10^{-4}$ & $-1.18\cdot10^{-4}$ & 
        $-6.37\cdot10^{-5}$ \\
    $7.34^\circ$ &
        $1.96\cdot10^{-2}$ & $3.41\cdot10^{-3}$ & $-2.57\cdot10^{-5}$ & 
        $-9.25\cdot10^{-5}$ \\
    $9.43^\circ$ &
        $1.13\cdot10^{-2}$ & $2.55\cdot10^{-3}$ & $-1.42\cdot10^{-3}$ & 
        $-2.46\cdot10^{-4}$ \\
    \hline\hline
    $\theta$ && $\phi_1$ & $\phi_2$ & $\phi_3$ \\
    \hline\hline
    $6.00^\circ$ &&
        $3.56\cdot10^{-1}$ & $-2.09\cdot10^{-10}$ & $4.53\cdot10^{-1}$ \\
    $7.34^\circ$ &&
        $1.63\cdot10^{-1}$ & $1.11\cdot10^{0}$ & $2.43\cdot10^{-2}$ \\
    $9.43^\circ$ &&
        $6.83\cdot10^{-1}$ & $-1.37\cdot10^{-1}$ & $-1.23\cdot10^{0}$ \\
    \hline\hline
    $\theta$ && $t_1^\mathrm{SOC}\,(\mathrm{\delete{m}eV})$ & $t_2^\mathrm{SOC}\,(\mathrm{\delete{m}eV})$ & $t_3^\mathrm{SOC}\,(\mathrm{\delete{m}eV})$ \\
    \hline\hline
    $6.00^\circ$ &&
        $1.94\cdot10^{-4}$ & $-8.43\cdot10^{-6}$ & $4.45\cdot10^{-6}$ \\
    $7.34^\circ$ &&
        $1.97\cdot10^{-4}$ & $5.85\cdot10^{-6}$ & $-1.55\cdot10^{-5}$ \\
    $9.43^\circ$ &&
        $4.89\cdot10^{-4}$ & $2.13\cdot10^{-4}$ & $2.12\cdot10^{-4}$ \\
    \hline\hline
  \end{tabular}
\end{table}

{\it Functional renormalization group} --- We treat the interacting two-band, non-$SU(2)$ tight binding model on the triangular lattice using the fRG. This method smoothly interpolates the free action $S^{\Lambda=\infty}$ to the full, interacting action $S^{\Lambda=0}$. We employ a sharp frequency cutoff scheme in the fermionic propagator:
\begin{equation}
    G^\Lambda_{\sigma\sigma'}(ik_0, \bvec k) = \Theta(|ik_0|-\Lambda) G^0_{\sigma\sigma'}(ik_0,\bvec k),
\end{equation}
with $\hat{G}^0(ik_0,\bvec k) = (ik_0-\hat{H}^0(\bvec k))^{-1}$. Numerical treatment is rendered possible by approximating the infinite hierarchy of flow equations \cite{Metzner2012a,platt-hanke-thomale2013functional,salmhofer-honerkamp2001fermionic} and discarding all vertices that describe more than four-fermion interactions as well as setting the four-point vertex $\Gamma^{(4),\Lambda}(1,2,3,4)$ constant for all incoming and outgoing frequencies. As we are interested in static properties, we further neglect frequency dependencies on the two-point vertex (self-energy) and thus set $\Gamma^{(2),\Lambda}(1,2)\equiv0$. By these approximations, we arrive at the following flow equations of the four-point vertex:
\begin{widetext}
    \def\diffop{\frac{\mathrm{d}}{\mathrm{d}\Lambda}}%
    \newcommand{\ostuff}[4]{\sigma_{#1}\sigma_{#2}\sigma_{#3}\sigma_{#4}}%
\begin{align}
  \diffop\,\Gamma^{(4),\Lambda} &{}= \includegraphics[valign=c]{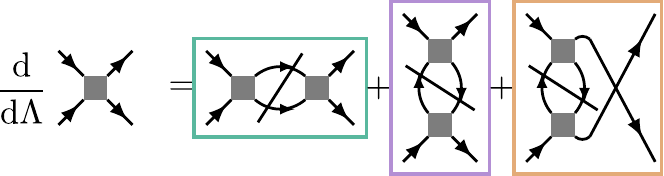} = \diffop\,\big(
    {\color{p-channel}P^\Lambda} +
    {\color{d-channel}D^\Lambda} +
    {\color{c-channel}C^\Lambda}
    \big) \,,\\
    \diffop P^\Lambda_{\ostuff1234}(\bvec q_P, \bvec k_P, \bvec k_P') &{}= 
        \frac12\sum_{\bvec k\ostuff{3'}{4'}{1'}{2'}}
    \Gamma^{P,\Lambda}_{\ostuff12{3'}{4'}}(\bvec q_P, \bvec k_P, \bvec k) \,\diffop L^{P,\Lambda}_{\ostuff{3'}{4'}{1'}{2'}}(\bvec q_P, \bvec k) \,
    \Gamma^{P,\Lambda}_{\ostuff{1'}{2'}34}(\bvec q_P, \bvec k, \bvec k_P') \,,
    \label{eq:frg-flow-P}
    \\
    \diffop D^\Lambda_{\ostuff1234}(\bvec q_D, \bvec k_D, \bvec k_D') &{}= 
        -\sum_{\bvec k\ostuff{3'}{4'}{1'}{2'}}
    \Gamma^{D,\Lambda}_{\ostuff12{3'}{4'}}(\bvec q_D, \bvec k_D, \bvec k) \,\diffop L^{D,\Lambda}_{\ostuff{3'}{4'}{1'}{2'}}(\bvec q_D, \bvec k) \,
    \Gamma^{D,\Lambda}_{\ostuff{1'}{2'}34}(\bvec q_D, \bvec k, \bvec k_D') \,,
    \label{eq:frg-flow-D}
    \\
    \diffop C^\Lambda_{\ostuff1234}(\bvec k_1, \bvec k_2, \bvec k_3) &{}=
    -\diffop D^\Lambda_{\ostuff1243}(\bvec k_1, \bvec k_2, \bvec k_1 + \bvec k_2 - \bvec k_3 )\,.
    \label{eq:frg-su2-equiv-c-d}
    \end{align}
\end{widetext}
The channel-projections read
{\def\ostuff#1#2#3#4{\sigma_{#1}\sigma_{#2}\sigma_{#3}\sigma_{#4}}%
\begin{align}
    \Gamma^{P,\Lambda}_{\ostuff1234}(\bvec q_P,\bvec k_P,\bvec k_P') &{}= \Gamma^{(4),\Lambda}_{\ostuff1234}(\bvec k_1,\bvec k_2,\bvec k_3) \,, \\
    \Gamma^{D,\Lambda}_{\ostuff1234}(\bvec q_D,\bvec k_D,\bvec k_D') &{}= \Gamma^{(4),\Lambda}_{\ostuff3124}(\bvec k_1,\bvec k_2,\bvec k_3) \,,
\end{align}}
with the momenta transformed as
\begin{align}
    \bvec q_P &{}= \bvec k_1 + \bvec k_2 \,, \quad
    \bvec k_P = \bvec k_1 \,, \quad
    \bvec k_P' = \bvec k_3 \,, \\
    \bvec q_D &{}= \bvec k_1 - \bvec k_3 \,, \quad
    \bvec k_D = \bvec k_1 \,, \quad
    \bvec k_D' = \bvec k_1 + \bvec k_2 - \bvec k_3 \,.
\end{align}
Note that the diagram contributions $P^\Lambda$ and $D^\Lambda$ are written in the respective channel-projected bases, whereas the relation from $P^\Lambda$ to $D^\Lambda$ (Eq.~\eqref{eq:frg-su2-equiv-c-d}) is given in the ordering where $1,2$ are ingoing and $3,4$ outgoing indices. The fermionic particle-particle and particle-hole loops have to be momentum-reordered as well with $\bvec k_2^P = \bvec q^P - \bvec k_1$, $\bvec k_2^D = \bvec q^D + \bvec k_1$  and read
\begin{widetext}
\def\diffop{\frac{\mathrm{d}}{\mathrm{d}\Lambda}}%
\def\ostuff#1#2#3#4{\sigma_{#1}\sigma_{#2}\sigma_{#3}\sigma_{#4}}%
\append{
\begin{align}
    \diffop L^{P,\Lambda}_{\ostuff1234}(\bvec q^P, \bvec k_1) &{}=
    \begin{multlined}[t]
    \frac1{2\pi}\,\sum_{b_1b_2}\,
    u_{\sigma_1b_1}^{\phantom{*}}(\bvec k_1) u_{\sigma_3b_1}^*(\bvec k_1)
    u_{\sigma_2b_2}^{\phantom{*}}(\bvec k_2^P) u_{\sigma_4b_2}^*(\bvec k_2^P) \,\times \\ \bigg[
    \frac1{(-i\Lambda-\epsilon_{b_1}(\bvec k_1))(i\Lambda-\epsilon_{b_2}(\bvec k_2^P))} + \frac1{(i\Lambda-\epsilon_{b_1}(\bvec k_1))(-i\Lambda-\epsilon_{b_2}(\bvec k_2^P))}
    \bigg] \,,
    \end{multlined} \\
    \diffop L^{D,\Lambda}_{\ostuff4123}(\bvec q^D, \bvec k_1) &{}=
    \begin{multlined}[t]
    \frac1{2\pi}\,\sum_{b_1b_2}\,
    u_{\sigma_1b_1}^{\phantom{*}}(\bvec k_1) u_{\sigma_3b_1}^*(\bvec k_1)
    u_{\sigma_2b_2}^{\phantom{*}}(\bvec k_2^D) u_{\sigma_4b_2}^*(\bvec k_2^D) \,\times
    \\ \bigg[
    \frac1{(i\Lambda-\epsilon_{b_1}(\bvec k_1))(i\Lambda-\epsilon_{b_2}(\bvec k_2^D))} + \frac1{(-i\Lambda-\epsilon_{b_1}(\bvec k_1))(-i\Lambda-\epsilon_{b_2}(\bvec k_2^D))}
    \bigg] \,,
    \end{multlined}
\end{align}
}
\end{widetext}
with $u_{\sigma b}(\bvec k)$ the Bloch functions of the non-interacting tight-binding Hamiltonian and $\epsilon_b(\bvec k)$ its dispersion.

We calculate the fRG flow on a regular $24\times24$ momentum mesh in the 2D primitive zone and take both the full spin- and momentum-structure of the four point vertex into account. The summation over $\bvec k$ in Eqs.~(\ref{eq:frg-flow-P},\ref{eq:frg-flow-D}) is carried out on a finer momentum mesh with $649$ points per coarse momentum point. The fine points are constructed to equally space out the Wigner-Seitz cells defined by the regular, coarse momentum mesh. To integrate the differential equation for $\Gamma^{(4),\Lambda}$, we employ an enhanced Euler scheme with adaptive step size chosen that the maximal step size can never be above $10\%$ of the current $\Lambda$. We consider the vertex diverged if its absolute maximal entry reaches $20$ times the system's bandwidth.

We analyze the instabilities in a two-fold procedure. First, we determine the divergent channel by inspecting whether particle-particle ($P^\Lambda$) or particle-hole ($D^\Lambda$) are the dominant contributions to make $\Gamma^{(4),\Lambda}$ diverge.

In the particle-particle case, we further investigate the superconducting instability by solving a linearized gap equation for $\Gamma^{(4),\Lambda}$:
\begin{widetext}
\begin{equation}
    \lambda\,\Delta_{\sigma_1\sigma_2}(\bvec k) = \sum_{\bvec k'\sigma_3\sigma_4\sigma_{1'}\sigma_{2'}} \Gamma^{P,\Lambda}_{\sigma_1\sigma_2\sigma_3\sigma_4}(\bvec q_P=0,\bvec k,\bvec k')\,
    L^{f,P,\Lambda}_{\sigma_3\sigma_4\sigma_{1'}\sigma_{2'}}(\bvec q^P=0,\bvec k')\,
    \Delta_{\sigma_{1'}\sigma_{2'}}(\bvec k')\,,
    \label{eq:frg-eigen}
\end{equation}
with the particle-particle (and particle-hole) fermi-loops given by
\begin{align}
    L^{f,P,\Lambda}_{\sigma_1\sigma_2\sigma_{3}\sigma_{4}}(\bvec q^P,\bvec k_1) &{}= \sum_{b_1b_2} \frac{
    u_{\sigma_1b_1}^{\phantom{*}}(\bvec k_1) u_{\sigma_3b_1}^*(\bvec k_1)
    u_{\sigma_2b_2}^{\phantom{*}}(\bvec k_2^P) u_{\sigma_4b_2}^*(\bvec k_2^P)\,
    \big[f\big(-\epsilon_{b_1}(\bvec k_1)/\Lambda\big) - f\big(\epsilon_{b_2}(\bvec k_2^P)/\Lambda\big)\big] }{
    \epsilon_{b_1}(\bvec k_1) + \epsilon_{b_2}(\bvec k_2^P)
    }\,,\\
    L^{f,D,\Lambda}_{\sigma_4\sigma_1\sigma_{2}\sigma_{3}}(\bvec q^D,\bvec k_1) &{}= \sum_{b_1b_2} \frac{
    u_{\sigma_1b_1}^{\phantom{*}}(\bvec k_1) u_{\sigma_3b_1}^*(\bvec k_1)
    u_{\sigma_2b_2}^{\phantom{*}}(\bvec k_2^D) u_{\sigma_4b_2}^*(\bvec k_2^D)\,
    \big[f\big(\epsilon_{b_1}(\bvec k_1)/\Lambda\big) - f\big(\epsilon_{b_2}(\bvec k_2^D)/\Lambda\big)\big] }{
    \epsilon_{b_1}(\bvec k_1) - \epsilon_{b_2}(\bvec k_2^D)
    }\,,
\end{align}
\end{widetext}
with the Fermi function $f(x)=(1+e^x)^{-1}$. As the eigenproblem in Eq.~\eqref{eq:frg-eigen} is non-hermitian and thus numerically highly unstable, we instead solve for the singular values and vectors of the matrix composed of $\Gamma^{P,\Lambda}$ and $L^{f,P,\Lambda}$:
\begin{equation}
    \hat{\Gamma}^{P,\Lambda}\hat{L}^{f,P,\Lambda} = \hat{U}\,\hat{\Sigma}\,\hat{V}^\dagger\,,
\end{equation}
where the right singular vectors $\hat V$ are the Fermi surface projected gap functions and the left singular vectors $\hat U$ lack the Fermi surface structure but instead show the gap's symmetry more clearly. We transform the gap function (i.e. leading singular vector) to singlet and triplet space using the following identity \cite{sigrist1991phenomenological,smidman2017superconductivity}:
\begin{equation}
    \hat\Delta(\bvec k) = i\,\delete{\hat\sigma_y}\big[ \hat\sigma_0\psi(\bvec k) + \hat{\bvec \sigma}\cdot\bvec d(\bvec k) \big]\,\append{\hat{\sigma}_y},
\end{equation}
with the vector of Pauli matrices $\hat{\bvec \sigma}$ and the identity matrix $\hat\sigma_0$.

The leading instability is doubly degenerate for a large part of the phase diagram. Thus, we compute the complex superposition of the two leading instabilities [cf. Eq.~\eqref{eq:frg-complex-sc}]. For $\vartheta\in(0,\pi)$ and $\varphi\in(0,\pi)$, we evaluate the free energy of the system in the superconducting state via
\begin{widetext}
\begin{multline}
    F^\Lambda(\vartheta,\varphi) = \frac1{N_{\bvec k}} \Bigg[
        \sum_{\bvec kb} f(E_b(\bvec k)/\Lambda)E_b(\bvec k) -
        \sum_{\bvec k\sigma_1\sigma_2\bvec k'\bvec\sigma_3\sigma_4} \Delta^{\vartheta,\varphi\,*}_{\sigma_1\sigma_2}(\bvec k) \big[\hat{\Gamma}^{P,\Lambda}\big]^{-1}_{\sigma_1\sigma_2\sigma_3\sigma_4}(\bvec k,\bvec k')
        \Delta^{\vartheta,\varphi}_{\sigma_3\sigma_4}(\bvec k') + \\
        \Lambda\sum_{\bvec kb} \bigg( f(E_b(\bvec k)/\Lambda)\log(f(E_b(\bvec k)/\Lambda))
        + f(-E_b(\bvec k)/\Lambda)\log(f(-E_b(\bvec k)/\Lambda)) \bigg) \Bigg]\,,
\end{multline}
\end{widetext}
where $E_b(\bvec k)$ is the dispersion of the Bogoljubov-de-Gennes Hamiltonian 
\begin{equation}
    \hat{H}^\mathrm{BdG}(\bvec k) = \begin{pmatrix}
    \hat{H}^0(\bvec k) & \hat{\Delta}^{\vartheta,\varphi}(\bvec k) \\
    \big[\hat{\Delta}^{\vartheta,\varphi}(\bvec k)\big]^\dagger & \big[-\hat{H}^{0}(-\bvec k)\big]^T
    \end{pmatrix}
\end{equation}
and $[\hat{\Gamma}^{P,\Lambda}]^{-1}$ is the pseudoinverse of the superconducting vertex as a matrix in $(\bvec k,\sigma_1,\sigma_2)$ and $(\bvec k',\sigma_3,\sigma_4)$. We evaluate  $F^\Lambda$ for $\Lambda=\Lambda_\mathrm{c}$. Next, we find the angles $\vartheta_0,\varphi_0$ at which the free energy is minimized. For the analysis of the topology in the superconducting state, we use the physically realized instability at $\vartheta_0$ and $\varphi_0$.

In the particle-hole case, we instead extract the spin/density susceptibility from the four-point vertex at the final scale given by
\begin{widetext}
\begin{equation}
    \chi^D_{\sigma_1\sigma_2\sigma_3\sigma_4}(\bvec q_D) = \sum_{\bvec k_D^{\phantom{\prime}}\bvec k_D^\prime\sigma_{1'}\sigma_{2'}\sigma_{3'}\sigma_{4'}}\,
    L^{f,D,\Lambda}_{\sigma_1\sigma_2\sigma_{1'}\sigma_{2'}}(\bvec q_D,\bvec k_D)\,
    \Gamma^{D,\Lambda}_{\sigma_{1'}\sigma_{2'}\sigma_{3'}\sigma_{4'}}(\bvec q_D, \bvec k_D, \bvec k_D')\,
    L^{f,D,\Lambda}_{\sigma_{3'}\sigma_{4'}\sigma_3\sigma_4}(\bvec q_D,\bvec k_D')\,.
\end{equation}
\end{widetext}
Subsequently, we transform the four-point susceptibility to physical channels \cite{scherer2018spin-orbit}:
\begin{equation}
    \chi^{ij}(\bvec q) = \sum_{\sigma_1\sigma_2\sigma_3\sigma_4} \sigma_i^{\sigma_1\sigma_3} \sigma_j^{\sigma_4\sigma_2}\,\chi^D_{\sigma_1\sigma_2\sigma_3\sigma_4}(\bvec q)\,.
\end{equation}
The density-density response is given by $\chi^{00}(\bvec q)$ and the spin response functions by $\chi^{xx}(\bvec q)$, $\chi^{yy}(\bvec q)$, $\chi^{zz}(\bvec q)$, $\chi^{xy}(\bvec q)$, $\chi^{xz}(\bvec q)$ and $\chi^{yz}(\bvec q)$.

\section*{Data Availability}
The raw data sets used for the presented analysis within the current study are available from the corresponding authors on reasonable request.

\section*{Code Availability}
The tailored developed codes used in this work can be provided from the corresponding author on reasonable request. Ab initio calculations are done with the code VASP (version 5.4.4).

\begin{acknowledgments}
We thank J.~Beyer and J.~Hauck for useful discussions on the generation and analysis of non-$SU(2)$ fRG results. This work is supported by the European Research Council (ERC-2015-AdG-694097), Grupos Consolidados (IT1249-19),  and SFB925. MC is supported by a startup grant from the University of Pennsylvania. AR is supported by the Flatiron Institute, a division of the Simons Foundation. We  acknowledge  funding by the Deutsche Forschungsgemeinschaft (DFG, German Research Foundation) under RTG 1995 and RTG 2247, within the Priority Program SPP 2244 ``2DMP'', under Germany's Excellence Strategy - Cluster of Excellence Matter and Light for Quantum Computing (ML4Q) EXC 2004/1 - 390534769 and - Cluster  of  Excellence and Advanced Imaging of Matter (AIM) EXC 2056 - 390715994. LX acknowledges the support from Distinguished Junior Fellowship program by the South Bay Interdisciplinary Science Center in the Songshan Lake Materials Laboratory and the Key-Area Research and Development Program of Guangdong Province of China (Grants No.2020B0101340001). We acknowledge computational resources provided by the Simons Foundation Flatiron Institute, the Max Planck Computing and Data Facility, RWTH Aachen University under project number rwth0716 and the Platform for Data-Driven Computational Materials Discovery of the Songshan Lake laboratory. This work was supported by the Max Planck-New York City Center for Nonequilibrium Quantum Phenomena.
\end{acknowledgments}


\end{document}